\documentclass[12pt,fleqn]{article}
\makeatletter
\@addtoreset{equation}{section}
\@addtoreset{table}{section}
\makeatother
\newcommand{\eref}[1]{(\ref{#1})}

\def\I{{\rm i}}
\def\D{{\rm d}}
\def\E{{\rm e}}
\newcommand{\vecvar}[1]{\mbox{\boldmath$#1$}}
\usepackage{amsmath,amssymb,array,fancybox,graphicx,psfrag}
\begin{document}
\vspace*{.5ex}
\noindent{\Large\bf A Super-Integrable Discretization of the Calogero Model}
\vspace{4ex}

\renewcommand{\thefootnote}{\fnsymbol{footnote}}

\noindent{\large\bf Hideaki Ujino}\footnote{Present address:
Gunma National College of Technology, Maebashi, Gunma 371--8530, Japan,
{\tt ujino@nat.gunma-ct.ac.jp}}

Centre de Recherches Math\'ematiques, Universit\'e de Montr\'eal,

Montreal, Quebec, Canada, H3C 3J7

\vspace{1ex}

\noindent{\large\bf Luc Vinet}

Department of Mathematics and Statistics and Department of Physics,

McGill University, Montreal, Quebec, Canada, H3A 2J5

{\tt luc.vinet@mcgill.ca}\vspace{1ex}

\noindent{\large\bf Haruo Yoshida}

National Astronomical Observatory of Japan,

Mitaka, Tokyo 181--8588, Japan

{\tt h.yoshida@nao.ac.jp}\vspace{1ex}

\paragraph{abstract:}A time-discretization that preserves
the super-integrability of the Calogero model is obtained by application
of the integrable time-discretization of the harmonic oscillator to
the projection method for the Calogero model with continuous time.
In particular, the difference equations of motion, which provide an explicit
scheme for time-integration, are explicitly presented for the two-body case.
Numerical results exhibit that the scheme conserves all the$(=3)$
conserved quantities of the (two-body) Calogero model with a precision
of the machine epsilon times the number of iterations.

\paragraph{Pacs:} 02.30.Ik, 02.60.Jh, 02.70.Bf

\setcounter{footnote}{0}
\renewcommand{\thefootnote}{\arabic{footnote}}
\section{Introduction}

Numerical analysis of dynamical systems has great importance 
and a wide variety of applications in science and engineering. 
The elaboration of schemes for numerical analysis has a long and 
continuous history of studies as well as a rich accumulation of techniques. 
Whenever one
applies numerical analysis to the equations of motion,
one must discretize the time-evolution of the dynamical system 
that is originally described by differential equations
because of the lack of the notion of infinity in numerical analysis.
This leads to difference equations, 
which do not usually describe the same dynamical system 
as the original one: time-discretization is typically accompanied 
by modification of the original system, which may cause a significant 
difference in the behavior of the solution from that of the original system, 
particularly after integration over a long period. 
Understanding and controlling such modifications are thus important 
in the quest for more accurate long-time integration in numerical analysis. 

The symplectic integration method 
(see, for instance, refs.~\cite{Hairer,Sanz,Yoshida93}),
or the symplectic integrator, is one of the time-discretizations that
was invented in such a quest. 
Given a Hamiltonian $H^{(0)}$, it is designed so that its 
one-step$(=\tau)$ time evolution gives the exact one-step time 
evolution of a modified Hamiltonian 
$\tilde{H}:=H^{(0)}+\tau H^{(1)}+\tau^2 H^{(2)}+\cdots$.
Since the modified Hamiltonian is conserved by the flow of the symplectic 
integrator, the fluctuation of the value of the original Hamiltonian  
$H^{(0)}$, i.e., the total energy of the system with continuous time, 
is bounded, which is far more favorable than unbounded increase or decrease 
of the total energy that one usually observes in other non-symplectic 
discretizations. This is the reason why the symplectic integration 
method shows better accuracy even after long-time integration.

However, even the symplectic integration method does not usually have 
modified constants of motion for all the constants of motion 
of a system, which often causes secular increase or decrease of 
the values of the constants of motion after long-time integration. 
For example, non-existence of the modified constants of motion is 
proved for the two super-integrable models discretized 
by the symplectic integrator: the two dimensional Harmonic oscillator 
with integer frequency ratio (except for the isotropic case) and the 
two-dimensional Kepler problem~\cite{Yoshida2001,Yoshida2002}.
The orbits generated by the symplectic integrator are not closed,
though those of the exact analytic solutions are closed, indeed.

Thus the following question naturally arises: are there any 
discretization schemes that preserve the (super-)integrability 
of (super-)integrable models? 
Actually, an extensive collection of the known integrable 
discretizations of integrable models is now available in the
monograph~\cite{Suris2003} that was published in recent years.
However, the super-integrable discretization of super integrable models
has not been studied that far yet. Quite recently,
an
affirmative answer to the question
on the super-integrable discretization
is shown for 
the Kepler problem for two and three dimensional 
cases~\cite{Minesaki2002,Minesaki2004}
, 
where 
the integrable discretization of the harmonic 
oscillator~\cite{Hirota2000} plays an essential role. 
The above super-integrable discretization conserves all the constants
of motion, i.e., the Hamiltonian, the angular momentum and the 
Runge--Lenz vector, and generates a sequence of discrete points on
the orbit of the exact analytic solution of the Kepler problem. And,
of course, the orbit with the eccentricity less than unity is closed.

The purpose of this paper is to present a super-integrable discretization 
of the Calogero model~\cite{Calogero1971,Sutherland1972},
\begin{equation}
  H:=\dfrac{1}{2}\sum_{i=1}^{N}(p_i^2+\omega^2 x_i^2)+\dfrac{1}{2}
  \sum_{\stackrel{\scriptstyle i,j=1}{i\neq j}}^{N}\dfrac{a^2}{(x_i-x_j)^2}.
  \label{eq:Calogero_Hamiltonian}
\end{equation}
The real-valued quantities $p_i$, $x_i$, $\omega$ and $a$ in the Hamiltonian
are the canonical momentum and coordinate of the $i$-th particle, 
the strength of the external harmonic confinement and the interaction 
parameter, respectively.
The Calogero model ($\omega\neq 0$) with $N$
degrees of freedom (corresponding to the $N$-body case) is 
maximally super-integrable in the sense that it has $2N-1$ constants of motion
which are independent of each other~\cite{Adler1977}. The super-integrable
structure of the Calogero model is built up by the Lax 
formulation~\cite{Lax1968,Moser1975}, with which the eigenvalue
problem of an oscillating Hermitian matrix~\cite{Perelomov1976} is
intrinsically involved. We shall present a discretization that preserves
the above super-integrable structure of the Calogero model for the general
$N$-body case. It gives, in particular, 
the explicit form of the difference equations of motion of the Calogero model
for the two-body case that conserves all the three constants of 
motion as
\begin{equation}
  \begin{split}
    \Delta_+ x_{1,n}=&\dfrac{1}{1+\frac{\omega^2\Delta t^2}{4}}
    \Bigl[p_{1,n}-\dfrac{\omega^2}{2}x_{1,n}\Delta t\Bigr]
    +\dfrac{1}{1+\frac{\omega^2\Delta t^2}{4}}
    \dfrac{2a}{x_{1,n}-x_{2,n}}M_{{\rm i},n}\Delta t\\
    & -2\Bigl[
    \dfrac{1-\frac{\omega^2\Delta t^2}{4}}{1+\frac{\omega^2\Delta t^2}{4}}
    x_{1,n}+\dfrac{1}{1+\frac{\omega^2\Delta t^2}{4}}p_{1,n}
    \Delta t\Bigr]M_{{\rm r},n}\Delta t\\
    &+\Bigl[
    \dfrac{1-\frac{\omega^2\Delta t^2}{4}}{1+\frac{\omega^2\Delta t^2}{4}}
    (x_{1,n}+x_{2,n})
    +\dfrac{1}{1+\frac{\omega^2\Delta t^2}{4}}(p_{1,n}+p_{2,n})\Delta t
    \Bigr]\bigl(M_{{\rm i},n}^2+M_{{\rm r},n}^2\Delta t^2\bigr)\Delta t,\\
    \Delta_+ p_{1,n}=&-\dfrac{\omega^2}{1+\frac{\omega^2\Delta t^2}{4}}
    \Bigl[x_{1,n}+\dfrac{1}{2}p_{1,n}\Delta t\Bigr]
    +\dfrac{1-\frac{\omega^2\Delta t^2}{4}}{1+\frac{\omega^2\Delta t^2}{4}}
    \dfrac{2a}{x_{1,n}-x_{2,n}}M_{{\rm i},n} \\
    & -2\Bigl[
    \dfrac{1-\frac{\omega^2\Delta t^2}{4}}{1+\frac{\omega^2\Delta t^2}{4}}
    p_{1,n}-\dfrac{\omega^2}{1+\frac{\omega^2\Delta t^2}{4}}x_{1,n}
    \Delta t\Bigr]M_{{\rm r},n}\Delta t\\
    &+\Bigl[
    \dfrac{1-\frac{\omega^2\Delta t^2}{4}}{1+\frac{\omega^2\Delta t^2}{4}}
    (p_{1,n}+p_{2,n})
    -\dfrac{\omega^2}{1+\frac{\omega^2\Delta t^2}{4}}(x_{1,n}+x_{2,n})\Delta t
    \Bigr]\bigl(M_{{\rm i},n}^2+M_{{\rm r},n}^2\Delta t^2\bigr)\Delta t,\\
    \Delta_+ x_{2,n}=&\Delta_+ x_{1,n}\Bigl|_{1\leftrightarrow 2},\quad
    \Delta_+ p_{2,n}=\Delta_+ p_{1,n}\Bigl|_{1\leftrightarrow 2},
  \end{split}
  \label{eq:difference_equation}
\end{equation}
with
\begin{align*}
    M_{{\rm i},n}:=&\dfrac{a}{\sqrt{4a^2\Delta t^2+Y_n^2}},\quad
    M_{{\rm r},n}:=
    \dfrac{2a^2}{4a^2\Delta t^2+Y_n^2+Y_n\sqrt{4a^2\Delta t^2+Y_n^2}},\\
    Y_n:=&\Bigl[1-\frac{\omega^2\Delta t^2}{4}\Bigr](x_{1,n}-x_{2,n})^2
    +(p_{1,n}-p_{2,n})(x_{1,n}-x_{2,n})\Delta t,
\end{align*}
where $x_{i,n}$, $p_{i,n}$, $i=1,2$, are the coordinate and the momentum of
the $i$-th particle at the $n$-th discrete time.
The symbol $\Delta_+$ denotes the advanced time-difference defined by
\begin{equation}
  \Delta_+ A_n:=\dfrac{A_{n+1}-A_n}{\Delta t},
  \label{eq:advanced_difference}
\end{equation}
for an arbitrary variable $A_n$ (e.g., $A_n=x_{i,n}$, $p_{i,n}$) 
with the discrete time $n$.

The paper is organized as follows. In section~\ref{sec:projection}, we
present a brief summary of the projection method~\cite{Perelomov1976} 
(for review, see refs.~\cite{Perelomov1981,Perelomov1990,Hoppe1992}, 
for example), which gives a solution
to the initial value problem of the Calogero model with continuous time.
The Lax equations for the Calogero model provide a map from the
Calogero model into the matrix-valued harmonic oscillator
and also play an essential role in our discretization.
Applying the integrable discretization for the harmonic 
oscillator~\cite{Hirota2000},
we discretize the projection method in section~\ref{sec:discretization}.
A discrete analogue of the Lax equations, which we call the dLax equations,
is derived as a natural consequence of the discretization.
In section~\ref{sec:DEM}, we present the explicit forms of the dLax equations
of the Calogero model for the two-body case, which are equivalent to
the difference equations of motion~\eref{eq:difference_equation}.
They provide an explicit scheme for the time-integration of the model.
Numerical results obtained by our integrable discretization 
as well as by two other discretization schemes, 
namely, the symplectic Euler and energy conservation methods,
are also presented.
Section~\ref{sec:remarks} is dedicated to the summary and concluding remarks.

\section{Projection Method}\label{sec:projection}

In terms of the Lax pair for the Calogero--Moser model~\cite{Moser1975},
\begin{equation}
  \begin{split}
      & L_{ij}(t):=p_{i}(t)\delta_{ij}
      +\dfrac{{\rm i}a}{x_i(t)-x_j(t)}(1-\delta_{ij}),\\
      & M_{ij}(t):=\sum_{k(\neq i)}
      \dfrac{{\rm i}a}{\bigl(x_i(t)-x_k(t)\bigr)^2}\delta_{ij}
      -\dfrac{{\rm i}a}{\bigl(x_i(t)-x_j(t)\bigr)^2}(1-\delta_{ij}), \qquad
        i,j=1,2,\ldots,N,
  \end{split}
  \label{eq:Lax_pair}
\end{equation}
as well as a diagonal matrix
\[
  D(t):={\rm diag}\bigl(x_1(t),x_2(t),\ldots,x_N(t)\bigr),
\]
the canonical equation of motion for the Calogero 
model~\eref{eq:Calogero_Hamiltonian}
can be cast into the Lax form,
\begin{align*}
  & \dfrac{\D L}{\D t}=[L,M]-\omega^2 D,\\
  & \dfrac{\D D}{\D t}=[D,M]+L,
\end{align*}
where $[A,B]:=AB-BA$, or equivalently,
\begin{equation}
  \dfrac{\D L^\pm}{\D t}=[L^\pm,M]\pm{\rm i}\omega L^\pm,
  \label{eq:Lax_equation}
\end{equation}
where
\begin{equation}
  L^\pm(t):=L(t)\pm{\rm i}\omega D(t). 
  \label{eq:L+-}
\end{equation}

The Lax equations~\eref{eq:Lax_equation} 
allow the following relation between the products 
of $L^+$ and $L^-$ and the Hamiltonian
\[
  \dfrac{\D}{\D t}\bigl((L^+)^{l_1}(L^-)^{m_1}\cdots\bigr)
  =[(L^+)^{l_1}(L^-)^{m_1}\cdots,M]
  +{\rm i}\omega\bigr(l_1+\cdots-(m_1+\cdots)\bigr)
  (L^+)^{l_1}(L^-)^{m_1}\cdots,
\]
for any nonnegative integers $l_1,m_1,\ldots$.
Iterated use of the above formula yields
\begin{align*}
  \dfrac{\D}{\D t}
  \Bigl(\prod_i{\rm Tr}\bigl(\prod_{j}^{\longrightarrow}
  (L^+)^{l_{i,j}}(L^-)^{m_{i,j}}\bigr)\Bigr)
  =&\sum_{i}\prod_{i\neq k}{\rm Tr}\bigl(\prod_j^{\longrightarrow}
  (L^+)^{l_{i,j}}(L^-)^{m_{i,j}}\bigr)
  {\rm Tr}\Bigl[\prod_{l}^{\longrightarrow}
  (L^+)^{l_{k,l}}(L^-)^{m_{k,l}},M\Bigr]\\
  & +\sum_{k,l}(l_{k,l}-m_{k,l})\prod_i{\rm Tr}\bigl(\prod_j^{\longrightarrow}
  (L^+)^{l_{i,j}}(L^-)^{m_{i,j}}\bigr)\\
  =&\sum_{k,l}(l_{k,l}-m_{k,l})\prod_i{\rm Tr}\bigl(\prod_j^{\longrightarrow}
  (L^+)^{l_{i,j}}(L^-)^{m_{i,j}}\bigr),
\end{align*}
where
\[
  \prod_j^{\longrightarrow}A_j:=A_1A_2\cdots.
\]
Thus the constants of motion of the Calogero model can be constructed by
taking the trace of any products of $L^+$ and $L^-$ of the following 
form,
\begin{equation}
  \prod_i{\rm Tr}\bigl(\prod_j^{\longrightarrow}
  (L^+)^{l_{i,j}}(L^-)^{m_{i,j}}\bigr),
  \quad\text{if }\sum_{i,j}l_{i,j}=\sum_{i,j}m_{i,j},
  \label{eq:constants_of_motion}
\end{equation}
which includes the constants of motion given 
in ref.~\cite{Adler1977}.
Considering the case $a=0$ where the matrices $L^\pm$ become diagonal,
one can confirm the quantities given 
eq.~\eref{eq:constants_of_motion} above include at least (and also at most)
$2N-1$ constants of motion that are independent of each other.
For example, one confirms
\begin{equation}
  \begin{split}
    & C_1:={\rm Tr}L^+{\rm Tr}L^-
    =\bigl(\sum_{i=1}^N p_i\bigr)^2+\omega^2\bigl(\sum_{i=1}^N x_i\bigr)^2,\\
    & I_1:={\rm Tr}L^+L^-=2H,\quad I_2={\rm Tr}(L^+)^2{\rm Tr}(L^-)^2,
  \end{split}
  \label{eq:sample_constants}
\end{equation}
are constants of motion of the Calogero model~\eref{eq:Calogero_Hamiltonian}
that are independent of each other.
Note that the relations satisfied by the Lax pairs~\cite{Barucchi1977,Ujino1996}
\begin{equation}
  \begin{split}
    & [L^+,L^-]=-2\I\omega[L,D]
    =-2\omega a(T-E), \quad T_{ij}:=1, \ E_{ij}:=\delta_{ij},\\
    & M{}^{\rm t}[1,\ldots,1]=0, \ [1,\ldots,1]M=0, \quad
    \bigl[\Leftrightarrow TM=MT=0\bigr]
  \end{split}
  \label{eq:[L+,L-]}
\end{equation}
particularly the relations in the second line of eq.~\eref{eq:[L+,L-]}
which we call the sum-to-zero property,
have played a crucial role in the quantum analogue of the Lax formulation.

The initial value problem of the Calogero model can be solved by
the projection method~\cite{Perelomov1976}, in which the Lax formulation 
presented above plays a crucial role. This method can be formulated
in an analogous way to the Dirac picture in the time-dependent
perturbation theory of quantum mechanics.
Let us introduce $L_{\rm D}^\pm$,
the $L^\pm$ matrix in the ``Dirac picture'' by
\begin{equation}
  L^\pm(t)=:\E^{\pm\I\omega t}L_{\rm D}^\pm(t),\quad
  L^\pm(0)=L_{\rm D}^\pm(0).
  \label{eq:Dirac_picture}
\end{equation}
Then the Lax equation~\eref{eq:Lax_equation} is rewritten as
\[
  \dfrac{\D L_{\rm D}^\pm}{\D t}=[L_{\rm D}^\pm,M],
\]
which has the same form as the Heisenberg equation in the Dirac picture
with the time-dependent perturbation 
$M(t):=M\bigl(x_1(t),x_2(t),\ldots,x_N(t)\bigr)$.

The above equation allows the formal solution as
\begin{equation}
  L_{\rm D}^\pm(t)=U^\dagger(0,t)L^\pm(0)U(0,t),
  \label{eq:formal_solution}
\end{equation}
where the time-evolution unitary matrix $U(t^\prime,t)$ is given by
the Dyson series of $M(t)$,
\begin{equation}
  U(t^\prime,t):=\sum_{k=0}^\infty\int_{t^\prime}^t\D t_k
  \int_{t^\prime}^{t_{k}}\D t_{k-1}
  \cdots\int_{t^\prime}^{t_2}\D t_1 M(t_1)\cdots M(t_{k-1})M(t_k),
  \label{eq:Dyson}
\end{equation}
which has the semigroup property:
\begin{equation}
  \begin{split}
    U(0,t)&:=U[\vecvar{x}(0),\vecvar{p}(0);a,\omega;t]\\
    &=U[\vecvar{x}(0),\vecvar{p}(0);a,\omega;t^\prime]
    U[\vecvar{x}(t^\prime),\vecvar{p}(t^\prime);a,\omega;t-t^\prime]
    =U(0,t^\prime)U(t^\prime,t),\\
    U^\dagger(0,t)&=U(t,0).
  \end{split}
  \label{eq:semi_group_c}
\end{equation}
Note that $U(t^\prime,t)$ has a constant eigenvector ${}^{\rm t}[1,\ldots,1]$
whose eigenvalue is unity,
\begin{equation}
  \begin{split}
    & U(t^\prime,t){}^{\rm t}[1,\ldots,1]={}^{\rm t}[1,\ldots,1], \ 
    [1,\ldots,1]U(t^\prime,t)=[1,\ldots,1] \\
    & \Leftrightarrow U(t^\prime,t)T=TU(t^\prime,t)=T.
  \end{split}
  \label{eq:eigenvector_Uc}
\end{equation}
This property of $U(t^\prime,t)$ is a consequence of the sum-to-zero
property of the matrix $M$ in eq.~\eref{eq:[L+,L-]}.

Substitution of the formal solution~\eref{eq:formal_solution} into
eqs.~\eref{eq:L+-} and \eref{eq:Dirac_picture} gives the following solution 
of the initial value problem of the Lax equation~\eref{eq:Lax_equation}:
\begin{equation}
  \begin{split}
    & D(t)=U^\dagger(0,t)\dfrac{\E^{\I\omega t}L^+(0)
    -\E^{-\I\omega t}L^-(0)}{2\I\omega}U(0,t),\\
    & L(t)=U^\dagger(0,t)\dfrac{\E^{\I\omega t}L^+(0)
    +\E^{-\I\omega t}L^-(0)}{2}U(0,t),
  \end{split}
  \label{eq:eigenvalue_problem}
\end{equation}
which means that the eigenvalues of the time-dependent Hermitian
matrix $\dfrac{\E^{\I\omega t}L^+(0)-\E^{-\I\omega t}L^-(0)}{2\I\omega}$
give the solution of the initial value problem of the Calogero model.
The time-evolution unitary matrix $U(0,t)$, which has been formally 
introduced as the Dyson series of $M(t)$, is given here as the 
diagonalizing matrix.

The unitary matrix $U(0,t)$ provides a map of the Calogero model into
the matrix-valued harmonic oscillator. Let us introduce the matrices
\begin{equation}
  \begin{split}
    & X^\pm(t):=U(0,t)L^\pm(t)U^\dagger(0,t)=P(t)\pm\I\omega Q(t),\\
    & P(t):=U(0,t)L(t)U^\dagger(0,t),\quad Q(t):=U(0,t)D(t)U^\dagger(0,t).
  \end{split}
  \label{eq:map_to_oscillator}
\end{equation}
Substituting $X^\pm$ into the Lax equation~\eref{eq:Lax_equation}, we have
\begin{equation}
  \dfrac{\D X^\pm}{\D t}=\pm\I\omega X^\pm,
  \label{eq:projected_equation1}
\end{equation}
which are equivalent to the equations of motion of the harmonic oscillator
\begin{equation}
  \dfrac{\D P}{\D t}=-\omega^2 Q, \quad \dfrac{\D Q}{\D t}=P.
  \label{eq:projected_equation2}
\end{equation}
Thus we confirm that the Lax equations of the Calogero 
model~\eref{eq:Lax_equation} can be mapped to the equations of motion 
of the matrix-valued harmonic oscillator, 
eqs.~\eref{eq:projected_equation1} and \eref{eq:projected_equation2}.
Though the Hermitian matrices $P(t)$ and $Q(t)$ can possess $2N^2$ parameters
for their initial values, their definitions~\eref{eq:map_to_oscillator} 
introduce the restriction in the initial values
\begin{equation}
  P(0)=L(0),\quad Q(0)=D(0),
  \label{eq:constraints}
\end{equation}
whose number $2N$ is the same as that of
the Calogero model. 
The solution of the initial value problem of the above equations of motion
is given by
\begin{equation}
  \begin{split}
    & Q(t)=D(0)\cos\omega t+\dfrac{L(0)}{\omega}\sin\omega t, \quad
      P(t)=L(0)\cos\omega t-\omega D(0)\sin\omega t.\\
    & \Bigl[\Leftrightarrow\ X^\pm(t)=\E^{\pm\I\omega t}L^{\pm}(0)\Bigr]
  \end{split}
  \label{eq:projected_solution}
\end{equation}
Substitution of the above solution~\eref{eq:projected_solution}
into the definition of $P(t)$ and $Q(t)$ in eq.~\eref{eq:map_to_oscillator}
reproduces the solution of the initial value problem of the Calogero 
model~\eref{eq:eigenvalue_problem}. In particular, 
the coordinates of the Calogero model is given by the eigenvalues
of the matrix-valued harmonic oscillator $Q(t)$ in 
eq.~\eref{eq:projected_solution}.
That is the essence of the projection method.

The restrictions on the initial values \eref{eq:constraints} can be 
explained in terms of the constraints on the variables $Q(t)$ and $P(t)$.
From eqs.~\eref{eq:[L+,L-]} and \eref{eq:map_to_oscillator},
one can derive
\begin{align*}
  [Q(t),P(t)] &= U(0,t)[D(t),L(t)]U^\dagger(0,t) \\
  & = \I a U(0,t)(T-E)U^\dagger(0,t).
\end{align*}
By use of the property of
the time-evolution unitary matrix~\eref{eq:eigenvector_Uc},
one obtains
\begin{equation}
  [Q,P]=\I a(T-E),
  \label{eq:nontrivial_constraints}
\end{equation}
which poses $N(N-1)$ constraints on the
``unconstrained'' variables $Q(t)$ and $P(t)$ given by 
two Hermitian matrices whose off-diagonal elements 
are pure imaginaries, i.e., 
\begin{equation}
  \begin{split}
    & Q_{ij}(t):=q_{ii}(t)\delta_{ij}+\I q_{ij}(t),\quad
    P_{ij}(t):=p_{ii}(t)\delta_{ij}+\I p_{ij}(t),\\
    & q_{ii}(t),\ q_{ij}(t),\ p_{ii}(t),\ p_{ij}(t)\in\mathbb{R},\quad
    q_{ij}(t)=-q_{ji}(t),\ p_{ij}(t)=-p_{ji}(t).
  \end{split}
  \label{eq:free-variables}
\end{equation}
The above restriction~\eref{eq:free-variables}
is consistent with the solution of the initial
value problem~\eref{eq:projected_solution}.
The number of the independent variables in $Q(t)$ and $P(t)$ given by
eq.~\eref{eq:free-variables} is $N(N+1)$. One thus reproduces the
number of the initial values $2N$ as 
the degrees of freedom of the constrained system, $N(N+1)-N(N-1)=2N$.
The transformation~\eref{eq:map_to_oscillator} thus should be interpreted
as a map of the Calogero model into the harmonic 
oscillator~\eref{eq:projected_equation2} of the Hermitian matrix 
given by eq.~\eref{eq:free-variables} with the
constraints~\eref{eq:nontrivial_constraints}.

\section{Integrable Discretization}\label{sec:discretization}
As we have discussed in the previous section, the equations of motion 
(the Lax equation) of the Calogero model can be mapped to those of the
matrix-valued harmonic oscillator. We thus begin with the integrable
discretization of the matrix-valued harmonic oscillator, whose difference
equations of motion are given as
\begin{equation}
  \begin{split}
    & \Delta_+ Q_n=\dfrac{1}{2}(P_{n+1}+P_n),\\
    & \Delta_+ P_n=-\dfrac{1}{2}\omega^2(Q_{n+1}+Q_n),
  \end{split}
  \label{eq:dharmonic}
\end{equation}
where $Q_n$ and $P_n$ are Hermitian matrices whose initial values are
fixed as $Q_0=D(0)$ and $P_0=L(0)$ so as to relate them with the difference 
analogue of the Calogero model. The difference equations given above
have the same form as those for one-dimensional harmonic 
oscillator discretized by the energy conservation 
scheme~\cite{Hirota2000}, 
which is nothing but the implicit midpoint rule giving a symplectic
integration method of order 2~(see Theorem VI.3.4 in ref.~\cite{Hairer}).

As in eq.~\eref{eq:map_to_oscillator}, we introduce the new variables
\[
  X_n^\pm:=P_n\pm\I\omega Q_n.
\]
This brings the 
difference equations of motion into
\[
  \Delta_+ X_n^\pm=\pm\dfrac{1}{2}\I\omega(X_{n+1}^\pm+X_n^\pm),
\]
which is equivalent to
\begin{equation}
  \Bigl(1\mp\dfrac{1}{2}\I\omega\Delta t\Bigr)X_{n+1}^\pm
  =\Bigl(1\pm\dfrac{1}{2}\I\omega\Delta t\Bigr)X_n^\pm.
  \label{eq:dLax_in_X}
\end{equation}
Defining the rescaled time-step by
\begin{equation}
  \Delta\tau:=\dfrac{2}{\omega}\arctan\dfrac{\omega\Delta t}{2},
  \label{eq:time_rescale}
\end{equation}
the recursion relation~\eref{eq:dLax_in_X} is rewritten as
\[
  X_{n+1}^\pm=\dfrac{1\pm\frac{\I\omega\Delta t}{2}}
  {1\mp\frac{\I\omega\Delta t}{2}}X_n^\pm
  =\E^{\pm\I\omega\Delta\tau}X_n^\pm.
\]
Thus the solution of the initial value problem of the discrete harmonic 
oscillator is
\[
  X_n^\pm=\E^{\pm\I n \omega\Delta\tau}X_0^\pm,
\]
or
\begin{equation}
  \begin{split}
    & Q_n=D(0)\cos n\omega\Delta\tau
    +\dfrac{1}{\omega}L(0)\sin n\omega\Delta\tau,\\
    & P_n=L(0)\cos n\omega\Delta\tau-\omega D(0)\sin n\omega\Delta\tau,
  \end{split}
  \label{eq:solution_of_dharmonic}
\end{equation}
in terms of $Q_n$ and $P_n$. The above solution~\eref{eq:solution_of_dharmonic}
is exactly the same as that for the harmonic oscillator with the continuous
time~\eref{eq:projected_solution} up to time-rescale~\eref{eq:time_rescale}.

The recursion relation~\eref{eq:dLax_in_X} provides us with an efficient way
to construct the constants of motion of the discrete harmonic
oscillator. Consider an arbitrary product of $X_{n+1}^+$ and $X_{n+1}^-$ like
$(X_{n+1}^+)^{l_1}(X_{n+1}^-)^{m_1}\cdots$, for $l_1,m_1=0,1,2\cdots$, 
and reverse the time for one discrete time-step using the recursion
relation~\eref{eq:dLax_in_X}. Then one obtains
\begin{equation}
  \begin{split}
    (X_{n+1}^+)^{l_1}(X_{n+1}^-)^{m_1}\cdots 
    & = \E^{\I\omega l_1\Delta\tau}(X_{n}^+)^{l_1}
    \E^{-\I\omega m_1\Delta\tau}(X_{n}^-)^{m_1}\cdots \\
    & =\E^{\I\omega (l_1+\cdots-(m_1+\cdots))\Delta\tau}
    (X_{n}^+)^{l_1}(X_{n}^-)^{m_1}\cdots.
  \end{split}
  \label{eq:trick_in_X}
\end{equation}
Taking the trace of the above relation, one obtains
\[
  {\rm Tr}\bigl(\prod_j^{\longrightarrow}
  (X_{n+1}^+)^{l_j}(X_{n+1}^-)^{m_j}\bigr)
  =\exp\bigl(\I\omega\Delta\tau\sum_k(l_k-m_k)\bigr)
  {\rm Tr}\bigl(\prod_j^{\longrightarrow}
  (X_{n}^+)^{l_j}(X_{n}^-)^{m_j}\bigr).
\]
Thus one can construct 
constants of motion of the discrete harmonic oscillator in a way 
parallel to what is done for the Calogero model~\eref{eq:constants_of_motion} by
\[
  \prod_i{\rm Tr}\bigl(\prod_j^{\longrightarrow}
  (X_n^+)^{l_{i,j}}(X_n^-)^{m_{i,j}}\bigr),\qquad
  \text{when }\sum_{i,j}l_{i,j}=\sum_{i,j}m_{i,j}
\]
as well as the matrix-valued constants of motion by
\[
  (X_{n+1}^+)^{l_1}(X_{n+1}^-)^{m_1}\cdots 
  = (X_{n}^+)^{l_1}(X_{n}^-)^{m_1}\cdots,
\]
when $l_1+\cdots=m_1+\cdots$.
In particular, the following quantity, 
\begin{equation}
  [X_n^+,X_n^-]=-2\omega a(T-E) \ \bigl[=[X_0^+,X_0^-]\bigr]
  \label{eq:conserved_in_X}
\end{equation}
is conserved since it is a special case of the
above matrix-valued constants of motion. Note that the constants of motion
of the matrix-valued harmonic oscillator with continuous time 
can be given by the same formulas.

As we have confirmed, the solution of the harmonic 
oscillator with discrete time~\eref{eq:solution_of_dharmonic} and 
that with the continuous time~\eref{eq:projected_solution} agree
up to time-rescale~\eref{eq:time_rescale}. Thus the eigenvalue of 
$Q_n$ must trace the same trajectory of the Calogero
model with continuous time. We shall discuss it more in detail.

Since the relations
$Q_n=Q(n\Delta\tau)$ and $P_n=P(n\Delta\tau)$ hold
as a consequence of eqs.~\eref{eq:projected_solution},
\eref{eq:time_rescale} and \eref{eq:solution_of_dharmonic},
we also have analogous relations with \eref{eq:map_to_oscillator}
for $Q_n$ and $P_n$,
\begin{equation}
  \begin{split}
    & D_n  =U^\dagger_n Q_n U_n = D(n\Delta\tau),\  
    L_n=U^\dagger_n P_n U_n = L(n\Delta\tau), \\
    & (D_n)_{ij}:=x_{i,n}\delta_{ij},\ 
    (L_n)_{ij}:=p_{i,n}\delta_{ij}
    +\dfrac{\I a}{x_{i,n}-x_{j,n}}(1-\delta_{ij}),
  \end{split}
  \label{eq:map_from_dharmonic}
\end{equation}
where the unitary matrix $U_n$ is also
given by the corresponding matrix in the theory for the model with 
continuous time:
\begin{equation}
  U_n=U(0,n\Delta\tau).
  \label{eq:d_diagonalizer}
\end{equation}
Introduce $L^\pm_n$ in analogy with $L^\pm$ in eq.~\eref{eq:L+-}, i.e.,
\begin{equation}
  L^\pm_n:=L_n\pm\I\omega D_n=U_n^\dagger X_n^\pm U_n
  \label{eq:dL+-}
\end{equation}
and substitute it into the recursion relation~\eref{eq:dLax_in_X}. 
Then one obtains
\[
  \Bigl(1\mp\dfrac{1}{2}\I\omega\Delta t\Bigr)
  U_{n+1}L^\pm_{n+1}U^\dagger_{n+1}
  =\Bigl(1\pm\dfrac{1}{2}\I\omega\Delta t\Bigr)
  U_nL^\pm_n U^\dagger_n.
\]
By multiplying with $U_n^\dagger$ on the left and $U_{n+1}$ on the right,
one obtains a recursion relation of the matrices $L_n^\pm$
\begin{equation}
  \begin{split}
    & \Bigl(1\mp\dfrac{1}{2}\I\omega\Delta t\Bigr)S_n L_{n+1}^\pm
      =\Bigl(1\pm\dfrac{1}{2}\I\omega\Delta t\Bigr)L_n^\pm S_n, \\
    & S_n:=U^\dagger_{n}U_{n+1},
  \end{split}
  \label{eq:dLax}
\end{equation}
which we shall call the discrete Lax equations, or in short, the dLax equations.
They will play an essential role in the construction of the constants of motion
for the discrete time model. From
the definition of the unitary matrix $S_n$ and using the semigroup property
of the time evolution unitary matrix $U(t^\prime,t)$~\eref{eq:semi_group_c},
one obtains
\begin{equation}
  \begin{split}
    & S_n:=U^\dagger_{n}U_{n+1}=U(n\Delta\tau,(n+1)\Delta\tau)
    =S_n[\vecvar{x}_n,\vecvar{p}_n;a,\omega;\Delta t],\\
    & U_n=S_1S_2\cdots S_{n-1}, \quad U_0=E.
  \end{split}
  \label{eq:S_in_U}
\end{equation}
This indicates that $S_n$ is the one-step time-evolution matrix.
We should note that 
{\it the discrete inhomogeneous Lax's equation} 
introduced for the Calogero--Moser model 
(the Hamiltonian~\eref{eq:Calogero_Hamiltonian} with $\omega=0$) 
in ref.~\cite{Nijhoff1994} inspired us with the above recursion 
equation~\eref{eq:dLax}.
However, the explicit forms of the Lax pair and the derivation of 
the recursion relation of this paper are different from those
in ref.~\cite{Nijhoff1994}.

The recursion relation~\eref{eq:dLax} can be interpreted as the discrete
time analogue of the Lax equation of the Calogero model because of the
following two reasons. The first reason is that the dLax 
equations~\eref{eq:dLax} reduces to the Lax equations of the Calogero 
model~\eref{eq:Lax_equation} in the continuous time limit, 
$\Delta t\rightarrow 0$. The other reason is that the dLax equations
also conserve the constants of motion of the Calogero model with
continuous time. We shall discuss them more in detail.

From the Taylor expansions of 
$L_{n+1}=L((n+1)\Delta\tau)$ and 
$S_{n}=U(t^\prime=n\Delta\tau,t=(n+1)\Delta\tau)$
together with the expression in the formal Dyson series~\eref{eq:Dyson}
at $t=n\Delta\tau$, one obtains
\begin{equation}
  \begin{split}
  & L_{n+1}^\pm\sim L_n^\pm+\dfrac{\D L_n^\pm}{\D t}\Delta t+O(\Delta t^2),\\
  & S_n\sim E+M_n\Delta t +O(\Delta t^2), \quad
  M_n:=M(n\Delta\tau).
  \end{split}
  \label{eq:Taylor}
\end{equation}
Substitution of the above expressions into the dLax equations~\eref{eq:dLax}
yields
\[
  (1\mp\dfrac{1}{2}\I\omega\Delta t)(E+M_n\Delta t)(L_n^\pm
  +\dfrac{\D L_n^\pm}{\D t}\Delta t)
  =(1\pm\dfrac{1}{2}\I\omega\Delta t)L_n^\pm(E+M_n\Delta t).
\]
Dividing the above relation by $\Delta t$ and taking the limit 
$\Delta t\rightarrow 0$, one gets
\[
  \dfrac{\D L_n^\pm}{\D t}=\bigl[L_n^\pm,M_n\bigr]
  \pm\I\omega L_n^\pm,
\]
which is nothing but the Lax equation of the Calogero model with
continuous time.

In a way parallel to the construction of the constants of motion
of the discrete harmonic oscillator, one can construct the constants 
of motion of the dLax equations~\eref{eq:dLax}. Using the
definition~\eref{eq:dL+-} of $L_n^\pm$ in eq.~\eref{eq:trick_in_X}
and multiplying by $U_n^\dagger$ and $U_{n+1}$ respectively
from the left and the right, one gets
\begin{align*}
  S_n(L_{n+1}^+)^{l_1}(L_{n+1}^-)^{m_1}\cdots 
  & = \E^{\I\omega l_1\Delta\tau}(L_{n}^+)^{l_1}
  \E^{-\I\omega m_1\Delta\tau}(L_{n}^-)^{m_1}\cdots S_n \\
  & =\E^{\I\omega (l_1+\cdots-(m_1+\cdots))\Delta\tau}
  (L_{n}^+)^{l_1}(L_{n}^-)^{m_1}\cdots S_n.
\end{align*}
Thus one has
\[
  (L_{n+1}^+)^{l_1}(L_{n+1}^-)^{m_1}\cdots 
  = \E^{\I\omega (l_1+\cdots-(m_1+\cdots))\Delta\tau}
  S_n^\dagger(L_{n}^+)^{l_1}(L_{n}^-)^{m_1}\cdots S_n.
\]
Since the trace of an arbitrary product 
of matrices is invariant under any cyclic change of the order of the matrices,
one has
\[
  {\rm Tr}\bigl(\prod_j^{\longrightarrow}
  (L_{n+1}^+)^{l_j}(L_{n+1}^-)^{m_j}\bigr)
  =\exp\bigl(\I\omega\Delta\tau\sum_k(l_k-m_k)\bigr)
  {\rm Tr}\bigl(\prod_j^{\longrightarrow}
  (L_{n}^+)^{l_j}(L_{n}^-)^{m_j}\bigr),
\]
which leads to
\begin{align*}
  \prod_i{\rm Tr}\bigl(\prod_{j}^{\longrightarrow}
  (L_{n+1}^+)^{l_{i,j}}(L_{n+1}^-)^{m_{i,j}}\bigr)
  =\exp\bigl(\I\omega\Delta\tau\sum_{k,l}(l_{k,l}-m_{k,l})\bigr)
  \prod_i
  {\rm Tr}\bigl(\prod_j^{\longrightarrow}
  (L_{n}^+)^{l_{i,j}}(L_{n}^-)^{m_{i,j}}\bigr).
\end{align*}
One thus concludes that 
\begin{equation}
  \prod_i
  {\rm Tr}\bigl(\prod_j^{\longrightarrow}
  (L_{n}^+)^{l_{i,j}}(L_{n}^-)^{m_{i,j}}\bigr)
  \label{eq:d_constants_of_motion}
\end{equation}
gives the constant of motion of the dLax equations~\eref{eq:dLax}
when
\[
  \sum_{i,j}l_{i,j}=\sum_{i,j}m_{i,j}.
\]
This is in complete agreement with the situation for 
the Calogero model with continuous time~\eref{eq:constants_of_motion}.

Since $L_n^\pm$ should have the same form as $L^\pm$ for the continuous
time model, the commutator between $L_n^+$ and $L_n^-$ also should be
a constant matrix as in eq.~\eref{eq:[L+,L-]}. In the continuous
time theory, this constant matrix is associated with the nontrivial
constraints of the matrix-valued harmonic 
oscillator~\eref{eq:nontrivial_constraints}. One can observe the
same situations in our discrete time model~\eref{eq:dLax}.
Substitution of eq.~\eref{eq:dL+-} into $[L_n^+,L_n^-]$ with the help of
eq.~\eref{eq:conserved_in_X} yields
\begin{align*}
  [L_n^+,L_n^-] &= U^\dagger_n[X_n^+,X_n^-]U_n\\
  &= U^\dagger_n\bigl(-2\omega a(T-E)\bigr)U_n.
\end{align*}
Since $U_n$ and also $S_n$ are made from the time-evolution unitary 
matrix $U(t^\prime,t)$ as in eqs.~\eref{eq:d_diagonalizer}
and \eref{eq:S_in_U}, they also have a constant eigenvector 
${}^{\rm t}[1,\ldots,1]$ whose eigenvalues are unity,
\begin{equation}
  \begin{split}
    & U_n{}^{\rm t}[1,\ldots,1]={}^{\rm t}[1,\ldots,1], \ 
    [1,\ldots,1]U_n=[1,\ldots,1]\Leftrightarrow U_nT=TU_n=T,\\
    & S_n{}^{\rm t}[1,\ldots,1]={}^{\rm t}[1,\ldots,1], \ 
    [1,\ldots,1]S_n=[1,\ldots,1]\Leftrightarrow S_nT=TS_n=T.
  \end{split}
  \label{eq:eigenvector_USd}
\end{equation}
Thus one immediately obtains
\begin{equation}
  [L_n^+,L_n^-]=-2\omega a(T-E).
  \label{eq:[L+,L-]d}
\end{equation}
using the property~\eref{eq:eigenvector_USd} of $U_n$.
The relation~\eref{eq:[L+,L-]d} is also equivalent to
\[
  [D_n,L_n]=\I a(T-E).
\]
Note that eq.~\eref{eq:map_from_dharmonic}
provides the most general form of $L_n$ that satisfies the above relation 
together with the diagonal matrix $(D_n)_{ij}=x_i\delta_{ij}$.

\section{Difference Equations of Motion}\label{sec:DEM}
Though we have developed an integrable discretization 
for the Calogero model in the previous section,
the difference equations of motion we have obtained is nevertheless
a formal expression. An explicit expression for $S_n$ is necessary
in order to obtain the dLax equations for the Calogero model explicitly.
To do this, we restrict the number of particles to two in the discussion
of the explicit form of the $S_n$ matrix.

From the dLax equations~\eref{eq:dLax}, one obtains
\[
  L_{n+1}\pm\I\omega D_{n+1}:=L_{n+1}^\pm
  =\dfrac{1\pm\frac{1}{2}\I\omega\Delta t}{1\mp\frac{1}{2}\I\omega\Delta t}
  S_n^\dagger L_n^\pm S_n.
\]
Solution of the above equations with respect to $D_{n+1}$ and $L_{n+1}$
are given by
\begin{equation}
  \begin{split}
    & D_{n+1}=\dfrac{1}{2\I\omega}S_n^\dagger\biggl[
      \dfrac{1+\frac{1}{2}\I\omega\Delta t}{1-\frac{1}{2}\I\omega\Delta t}L_n^+
      -\dfrac{1-\frac{1}{2}\I\omega\Delta t}{1+\frac{1}{2}\I\omega\Delta t}L_n^-
      \biggr]S_n, \\
    & L_{n+1}=\dfrac{1}{2}S_n^\dagger\biggl[
      \dfrac{1+\frac{1}{2}\I\omega\Delta t}{1-\frac{1}{2}\I\omega\Delta t}L_n^+
      +\dfrac{1-\frac{1}{2}\I\omega\Delta t}{1+\frac{1}{2}\I\omega\Delta t}L_n^-
      \biggr]S_n.
  \end{split}
  \label{eq:S_as_diagonalizer}
\end{equation}
Since $D_{n+1}$ is diagonal, $S_n$ diagonalizes
\[
  \dfrac{1+\frac{1}{2}\I\omega\Delta t}{1-\frac{1}{2}\I\omega\Delta t}L_n^+
  -\dfrac{1-\frac{1}{2}\I\omega\Delta t}{1+\frac{1}{2}\I\omega\Delta t}L_n^-.
\]
$S_n$ thus can be constructed from the eigenvectors 
$\vecvar{v}_{i,n}$, $i=1,\cdots,N$ of the above matrix.
Normalized eigenvectors $\vecvar{s}_{i,n}$ that constitute
$S_n=[\vecvar{s}_{1,n}\cdots\vecvar{s}_{N,n}]$ should have the following
form, 
\begin{equation}
  \vecvar{s}_{i,n}=\dfrac{\vecvar{v}_{i,n}}{\sum_{j=1}^N(\vecvar{v}_{i,n})_j}
  \Rightarrow \sum_{j=1}^N(\vecvar{s}_{i,n})_j=1
  \label{eq:normalization}
\end{equation}
so as to satisfy the condition~\eref{eq:eigenvector_USd}.
Ordering of the eigenvectors
$\vecvar{s}_{i,n}$ is uniquely determined by the Taylor expansion 
of $S_n$~\eref{eq:Taylor}, or equivalently
\begin{equation}
  \lim_{\Delta t\rightarrow 0}S_n=E.
  \label{eq:ordering}
\end{equation}
Considering the eqs.~\eref{eq:S_as_diagonalizer} \eref{eq:normalization}
and \eref{eq:ordering} all together, one confirms that
\begin{equation}
  \begin{split}
    S_n &= E+\dfrac{2a\I\Delta t}{Y_n-2a\I\Delta t+\sqrt{4a^2\Delta t^2+Y_n^2}}
    \left[\begin{array}{rr}
    1 & -1 \\
    -1 & 1
    \end{array}\right] \\
    &= E + (\I M_{{\rm i},n}\Delta t-M_{{\rm r},n}\Delta t^2)
    \left[\begin{array}{rr}
    1 & -1 \\
    -1 & 1
    \end{array}\right] \sim E+M_n\Delta t \quad (\Delta t\rightarrow 0)
  \end{split}
  \label{eq:explicit_S}
\end{equation}
gives the explicit form of the $S_n$ matrix of the dLax equations 
for the Calogero model for the two-body case. It is straightforward to verify
that the dLax equations~\eref{eq:dLax} with the 
above $S_n$ matrix~\eref{eq:explicit_S} yields the difference equations of
motion~\eref{eq:difference_equation} shown in the introduction.

In order to confirm how well the dLax equations~\eref{eq:dLax} with the 
$S_n$ matrix~\eref{eq:explicit_S} or
the difference equations of motion~\eref{eq:difference_equation}
of the Calogero model with discrete time
trace the behavior of that with continuous time, the explicit expression
of the analytic solution for the continuous-time model is of great help,
which is given by the eigenvalues of $Q(t)$ in eq.~\eref{eq:projected_solution}.
But there is another derivation.
Since $U(t^\prime,t)$ is related to $S_n$ 
according to eq.~\eref{eq:S_in_U} 
with $t^\prime:=n\Delta\tau$ and $t:=(n+1)\Delta\tau$,
one can obtain the explicit form of $U(t^\prime,t)$ for the two-body case
from that of $S_n$~\eref{eq:explicit_S}:
\begin{equation}
  \begin{split}
    & U(t^\prime,t)=E+
    \dfrac{\frac{2a\I\sin\omega(t-t^\prime)}{\omega}}
    {{\mathcal Y}-\frac{2a\I\sin\omega(t-t^\prime)}{\omega}
    +\sqrt{\frac{4a^2\sin^2\omega(t-t^\prime)}{\omega^2}+{\mathcal Y}^2}}
    \left[\begin{array}{rr}
    1 & -1 \\
    -1 & 1
    \end{array}\right], \\
    & {\mathcal Y}:=(x_1-x_2)^2\cos\omega(t-t^\prime)+(p_1-p_2)(x_1-x_2)
    \dfrac{\sin\omega(t-t^\prime)}{\omega}.
  \end{split}
  \label{eq:explicit_U}
\end{equation}
Substitution of the explicit forms of $U(0,t)$~\eref{eq:explicit_U} and
$L^\pm(0)$ for the two-body case into eq.~\eref{eq:eigenvalue_problem},
one obtains the explicit form of the analytic solution of the two-body
Calogero model with continuous time:
\begin{equation}
  \begin{split}
    x_1(t)=&\dfrac{1}{2}\Bigl[\bigl(x_{1,0}+x_{2,0}\bigr)\cos\omega t
    +\dfrac{1}{\omega}\bigl(p_{1,0}+p_{2,0}\bigr)\sin\omega t\Bigr] 
    \\
    +&\dfrac{1}{2(x_{1,0}-x_{2,0})}\sqrt{\frac{4a^2\sin^2\omega t}{\omega^2}
    +\Bigl[
      (x_{1,0}-x_{2,0})^2\cos\omega t
      +(p_{1,0}-p_{2,0})(x_{1,0}-x_{2,0})
    \frac{\sin\omega t}{\omega}
    \Bigr]^2}\\
    =:&x_1[\vecvar{x}_0,\vecvar{p}_0;a,\omega;t]\\
    p_1(t)=&\dfrac{1}{2}\Bigl[\bigl(p_{1,0}+p_{2,0}\bigr)\cos\omega t
    -\omega\bigl(x_{1,0}+x_{2,0}\bigr)\sin\omega t\Bigr] \\
    +&\dfrac{1}{2(x_{1,0}-x_{2,0})\sqrt{\frac{4a^2\sin^2\omega t}{\omega^2}
    +\Bigl[
      (x_{1,0}-x_{2,0})^2\cos\omega t
      +(p_{1,0}-p_{2,0})(x_{1,0}-x_{2,0})
    \frac{\sin\omega t}{\omega}
    \Bigr]^2}}\\
    & \times\biggl[\frac{2a^2\sin2\omega t}{\omega}+
    \Bigl[
      (x_{1,0}-x_{2,0})^2\cos\omega t
      +(p_{1,0}-p_{2,0})(x_{1,0}-x_{2,0})
    \frac{\sin\omega t}{\omega}
    \Bigr]\\
    &\times
    \Bigl[
      -\omega(x_{1,0}-x_{2,0})^2\sin\omega t
      +(p_{1,0}-p_{2,0})(x_{1,0}-x_{2,0})\cos\omega t
    \Bigr]
    \biggr]\\
    =:&p_1[\vecvar{x}_0,\vecvar{p}_0;a,\omega;t]\\
    x_2(t)=&x_1(t)\Bigr|_{1\leftrightarrow 2},\quad
    p_2(t)=p_1(t)\Bigr|_{1\leftrightarrow 2}
  \end{split}
  \label{eq:analytic_solution}
\end{equation}
The above expressions are used to numerically display the exact analytic 
results in the following figures. 
The initial values, the interaction parameter,
the strength of the external harmonic well are set at
$(x_1,p_1,x_2,p_2)|_{t=0}=(-4.00, 5.00, 2.00, 1.00)$,
$a = 3.00$ and $\omega = 0.314$ throughout the numerical calculation
in the following.

\begin{figure}[htb]
\psfrag{momenta}{momenta}
\psfrag{coordinates}{coordinates}
\psfrag{p1}{particle 1}
\psfrag{p2}{particle 2}
\psfrag{analytic}{analytic}
\psfrag{discrete}{discrete}
\psfrag{time}{time $t$}
\psfrag{4}{4}
\psfrag{2}{2}
\psfrag{0}{0}
\psfrag{-2}{$-2$}
\psfrag{-4}{$-4$}
\psfrag{15}{15}
\psfrag{10}{10}
\psfrag{5}{5}
\psfrag{0}{0}
\psfrag{-5}{$-5$}
\psfrag{-10}{$-10$}
\psfrag{-15}{$-15$}
\psfrag{4810}{4810}
\psfrag{4820}{4820}
\psfrag{4830}{4830}
\psfrag{4840}{4840}
\begin{center}
  \includegraphics[width=100mm]{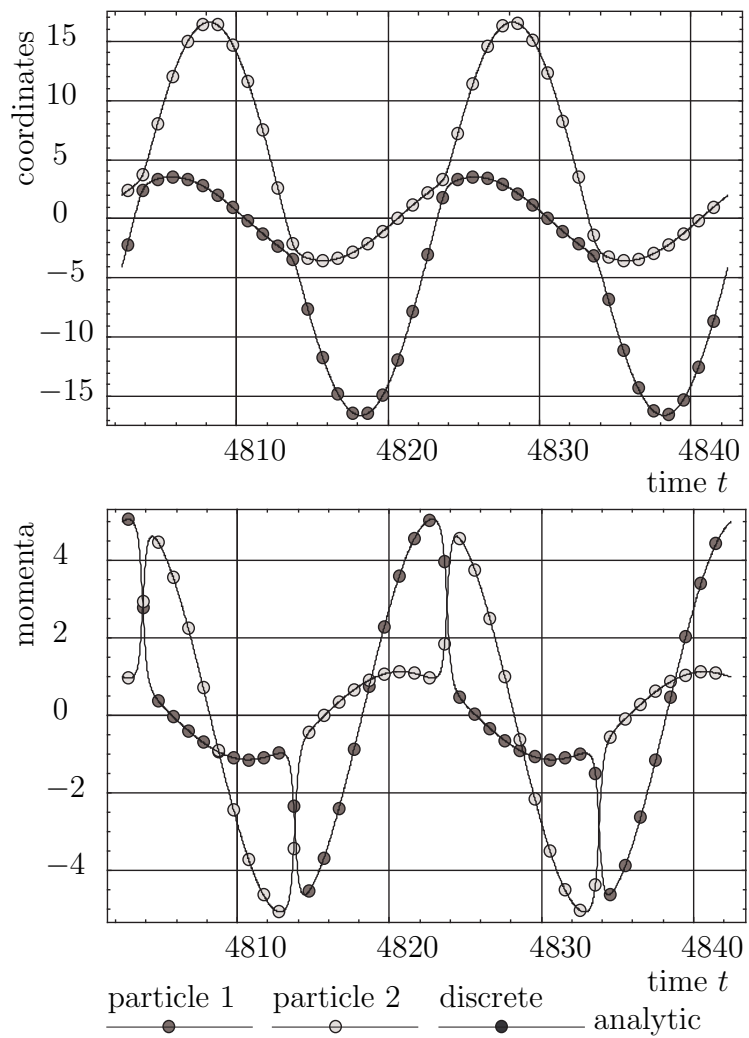}
\end{center}
\caption{The time evolution of the coordinates and the momenta 
generated by the analytic solution and the difference 
equations of motion~\eref{eq:difference_equation}.}
\label{fig:1}
\end{figure}

Figure~\ref{fig:1} presents the time evolution of the coordinates and
the momenta generated by the analytic solution and the difference 
equations of motion~\eref{eq:difference_equation} that gives the
super-integrable discretization.
The time-step in the super-integrable discretization is set at 
$\Delta t = 1.00$. The relation between the time $t$ and the number
of iterations $n$ is given by $n=t/\Delta\tau$ where $\Delta\tau=0.991903$.
The time interval $240\times\frac{2\pi}{\omega}=
4802\leq t\leq 4842=242\times\frac{2\pi}{\omega}$ corresponds
to $4841\leq n \leq 4881$ in terms of the number of iterations
for the discrete case. 
As has been confirmed by 
the correspondence between the matrices $L$ and $D$ for 
the super-integrable discretization and those for
the continuous time model~\eref{eq:map_from_dharmonic},
the solution of the discrete equations of 
motion~\eref{eq:difference_equation} gives a sequence of
canonical variables that are ``sampled'' from the orbit of the
exact analytic solution, i.e.,
\begin{equation}
  x_{i,n}=x_i(t=n\Delta\tau),\ p_{i,n}=p_i(t=n\Delta\tau).
  \label{eq:d-c_correspondence}
\end{equation}
Even though the coordinates and the momenta generated by 
the difference equations of motion~\eref{eq:difference_equation} are
computed through  sufficiently large numbers of 
iterations, they are in quite good agreement with those of the exact 
analytic solution, which gives a good numerical confirmation of the 
correspondence~\eref{eq:d-c_correspondence}.

One can confirm the precise agreement of the two solutions in a more
reliable manner by observing the relative errors of the constants of motion.
Using eq.~\eref{eq:constants_of_motion} 
(or eq.~\eref{eq:d_constants_of_motion}), we have already given three
constants of motion of the Calogero model~\eref{eq:sample_constants}.
Note that $C_1$ and $I_1$ correspond to the Hamiltonian
for the center of mass and the Calogero Hamiltonian, respectively.
Those three constants of motion give a set of independent conserved quantities.
But just for convenience, we introduce a set of slightly modified constants 
of motion:
\begin{equation}
  \begin{split}
    & C_1=(p_1+p_2)^2+\omega^2(x_1+x_2)^2,\\
    & C_2:=2I_1-C_1=(p_1-p_2)^2+\omega^2(x_1-x_2)^2+\dfrac{4a^2}{(x_1-x_2)^2},\\
    & C_3:=\dfrac{C_1^2-I_2}{4\omega^2}=(x_1p_2-x_2p_1)^2
    +\dfrac{2a^2(x_1^2+x_2^2)}{(x_1-x_2)^2}.
  \end{split}
  \label{eq:C1C2C3}
\end{equation}
Note that $C_2$ is the Hamiltonian for the relative coordinates and $C_3$
can be interpreted as a ``modified quadratic angular momentum.''
We use the above $C_i$'s, $i=1,2,3,$ for numerical calculation.

\begin{figure}
\psfrag{time}{time $t$}
\psfrag{relative errors}{relative errors}
\psfrag{0}{0}
\psfrag{1000}{1000}
\psfrag{2000}{2000}
\psfrag{3000}{3000}
\psfrag{4000}{4000}
\psfrag{5000}{5000}
\psfrag{1500}{${\scriptscriptstyle 1500}$}
\psfrag{1520}{}
\psfrag{1540}{}
\psfrag{1560}{}
\psfrag{1580}{}
\psfrag{1600}{${\scriptscriptstyle 1600}$}
\psfrag{3.2}{\hspace*{-1em}${\scriptscriptstyle 3.2\times 10^{-14}}$}
\psfrag{3}{${\hspace*{-1em}\scriptscriptstyle 3.0\times 10^{-14}}$}
\psfrag{2.8}{\hspace*{-1em}${\scriptscriptstyle 2.8\times 10^{-14}}$}
\psfrag{1.22}{\hspace*{-1.2em}${\scriptscriptstyle 1.22\times 10^{-13}}$}
\psfrag{1.25}{\hspace*{-1.2em}${\scriptscriptstyle 1.25\times 10^{-13}}$}
\psfrag{1.28}{\hspace*{-1.2em}${\scriptscriptstyle 1.28\times 10^{-13}}$}
\psfrag{1.76}{\hspace*{-1.2em}${\scriptscriptstyle 1.76\times 10^{-13}}$}
\psfrag{1.74}{\hspace*{-1.2em}${\scriptscriptstyle 1.74\times 10^{-13}}$}
\psfrag{1.72}{\hspace*{-1.2em}${\scriptscriptstyle 1.72\times 10^{-13}}$}
\psfrag{1.7}{\hspace*{-1.2em}${\scriptscriptstyle 1.70\times 10^{-13}}$}\psfrag{order}{$1$}
\psfrag{C1}{\tiny$C_1^{\rm si}$}
\psfrag{C2}{\tiny$C_2^{\rm si}$}
\psfrag{C3}{\tiny$C_3^{\rm si}$}
\begin{center}
  \includegraphics[width=120mm]{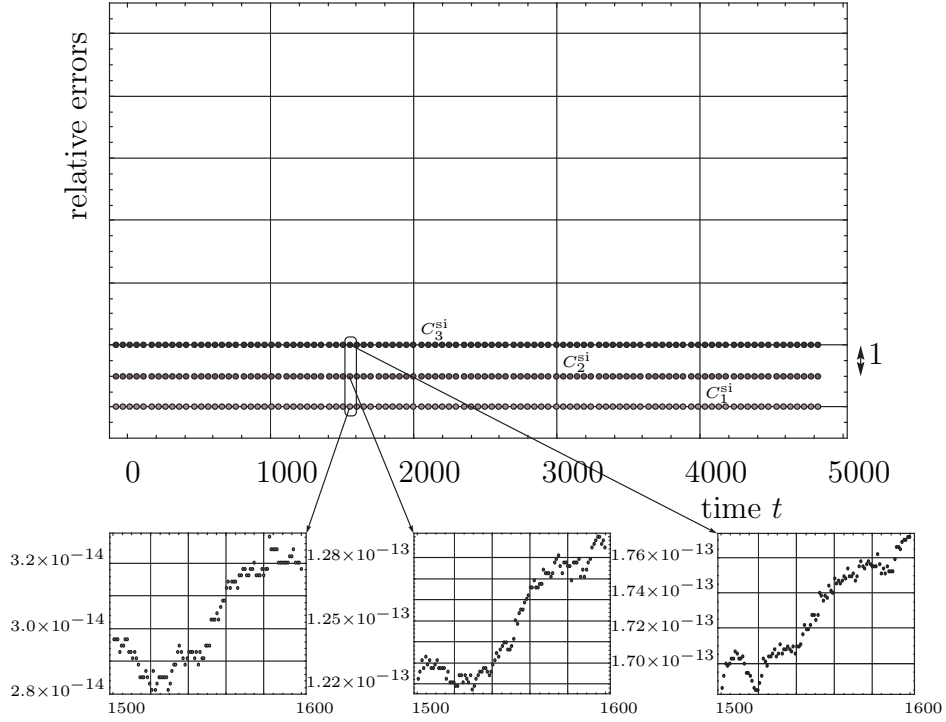}
\end{center}
\caption{The relative errors of the constants of motion
$C_i^{\rm si}:=C_{i,n}/C_{i,0}$, $n=t/\Delta\tau$,
of the discrete solutions given by the super-integrable
discretization~\eref{eq:difference_equation}.}
\label{fig:2}
\end{figure}

Figure~\ref{fig:2} presents the relative errors of the constants of motion
$C_i^{\rm si}:=C_{i,n}/C_{i,0}$, $n=t/\Delta\tau$ 
of the discrete solutions given by the super-integrable
discretization~\eref{eq:difference_equation}.
The offsets ($=(i-1)\times1.0$) are added for convenience 
of presentation.
Even though one sees the growth of errors caused by round-off errors
that is inevitable in any numerical analysis,
one confirms that all the constants of motion are conserved with a precision
of the machine epsilon times the number of iterations.
Thus one concludes that fig.~\ref{fig:2} presents 
a direct numerical verification of the fact that the difference 
equation of  motion~\eref{eq:difference_equation} preserves
the super-integrability of the Calogero model.

Let us compare the above results with
numerical demonstrations by two other discretizations.
The first one is the energy conservation scheme~(see pp.~159--161 in 
ref.~\cite{Hairer}),
\begin{equation}
  \begin{split}
    & \Delta_+ x_{i,n}=\dfrac{1}{2}(p_{i,n+1}+p_{i,n}),\\
    & \Delta_+ p_{i,n}=-\dfrac{1}{2}\omega^2(x_{i,n+1}+x_{i,n})
    +\sum_{\stackrel{\scriptstyle j=1}{j\neq i}}^{N}
    \dfrac{a^2\bigl((x_{i,n+1}-x_{j,n+1})+(x_{i,n}-x_{j,n})\bigr)}
    {(x_{i,n+1}-x_{j,n+1})^2(x_{i,n}-x_{j,n})^2},
  \end{split}
  \label{eq:energy_conservation_scheme}
\end{equation}
which keeps two constants of motion $C_1$ and $I_1$
(consequently, $C_2$) exactly for arbitrary number of particles $N$.
As one can see, the scheme~\eref{eq:energy_conservation_scheme} is
an implicit scheme, which involves numerical solution of
simultaneous algebraic equations.
The other scheme is the symplectic Euler 
method~(see Theorem VI.3.3 in ref.~\cite{Hairer}),
\begin{equation}
  \begin{split}
    & \Delta_+x_{i,n}=p_{i,n}, \\
    & \Delta_+p_{i,n}=-\omega^2 x_{i,n+1}
    +\sum_{\stackrel{\scriptstyle j=1}{j\neq i}}^{N}
    \dfrac{2a^2}{(x_{i,n+1}-x_{j,n+1})^3},
  \end{split}
  \label{eq:symplectic_Euler}
\end{equation}
which is an explicit scheme. Note that $\Delta_+$ denotes the advanced
time-difference~\eref{eq:advanced_difference}.

\begin{figure}
\psfrag{time}{time $t$}
\psfrag{relative errors}{relative errors}
\psfrag{0}{0}
\psfrag{1000}{1000}
\psfrag{2000}{2000}
\psfrag{3000}{3000}
\psfrag{4000}{4000}
\psfrag{5000}{5000}
\psfrag{1500}{${\scriptscriptstyle 1500}$}
\psfrag{1520}{}
\psfrag{1540}{}
\psfrag{1560}{}
\psfrag{1580}{}
\psfrag{1600}{${\scriptscriptstyle 1600}$}
\psfrag{1.24}{\hspace*{-1.15em}${\scriptscriptstyle 1.24\times 10^{-12}}$}
\psfrag{1.22}{${\hspace*{-1.15em}\scriptscriptstyle 1.22\times 10^{-12}}$}
\psfrag{a}{\hspace*{-1.15em}${\scriptscriptstyle 1.20\times 10^{-12}}$}
\psfrag{1.18}{\hspace*{-1.15em}${\scriptscriptstyle 1.18\times 10^{-12}}$}
\psfrag{2}{\hspace*{-1em}${\scriptscriptstyle 2.0\times 10^{-13}}$}
\psfrag{1.8}{\hspace*{-1em}${\scriptscriptstyle 1.8\times 10^{-13}}$}
\psfrag{1.6}{\hspace*{-1em}${\scriptscriptstyle 1.6\times 10^{-13}}$}
\psfrag{b}{\hspace*{-1em}${\scriptscriptstyle 1.4\times 10^{-13}}$}
\psfrag{c}{\hspace*{-1em}${\scriptscriptstyle 1.2\times 10^{-13}}$}
\psfrag{1.4}{\hspace*{1em}${\scriptscriptstyle 1.4}$}
\psfrag{1.2}{\hspace*{1em}${\scriptscriptstyle 1.2}$}
\psfrag{1}{${\hspace*{1em}\scriptscriptstyle 1.0}$}
\psfrag{0.8}{\hspace*{1em}${\scriptscriptstyle 0.8}$}
\psfrag{0.6}{\hspace*{1em}${\scriptscriptstyle 0.6}$}
\psfrag{0.4}{\hspace*{1em}${\scriptscriptstyle 0.4}$}
\psfrag{order}{$1$}
\psfrag{C1}{\tiny$C_1^{\rm ec}$}
\psfrag{C2}{\tiny$C_2^{\rm ec}$}
\psfrag{C3}{\tiny$C_3^{\rm ec}$}
\begin{center}
  \includegraphics[width=120mm]{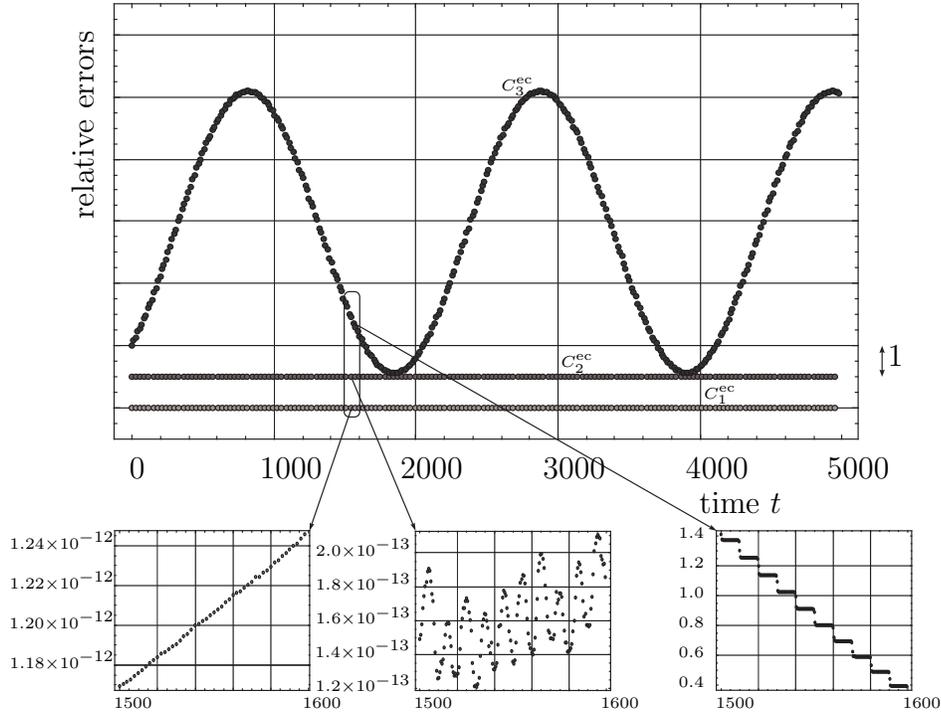}
\end{center}
\caption{The relative errors of the constants of motion,
$C_i^{\rm ec}:=C_{i,n}/C_{i,0}$, $n=t/\Delta t$, of 
the discrete solution given by the energy conservation 
scheme~\eref{eq:energy_conservation_scheme}.}
\label{fig:3}
\end{figure}

\begin{figure}
\psfrag{time}{time $t$}
\psfrag{relative errors}{relative errors}
\psfrag{z}{0}
\psfrag{1000}{1000}
\psfrag{2000}{2000}
\psfrag{3000}{3000}
\psfrag{4000}{4000}
\psfrag{5000}{5000}
\psfrag{1500}{${\scriptscriptstyle 1500}$}
\psfrag{1520}{}
\psfrag{1540}{}
\psfrag{1560}{}
\psfrag{1580}{}
\psfrag{1600}{${\scriptscriptstyle 1600}$}
\psfrag{0.02}{\hspace*{1em}${\scriptscriptstyle 0.02}$}
\psfrag{0.01}{\hspace*{1em}${\scriptscriptstyle 0.01}$}
\psfrag{0}{${\hspace*{1em}\scriptscriptstyle 0.00}$}
\psfrag{-0.01}{\hspace*{.5em}${\scriptscriptstyle -0.01}$}
\psfrag{-0.02}{\hspace*{.5em}${\scriptscriptstyle -0.02}$}
\psfrag{-0.03}{\hspace*{.5em}${\scriptscriptstyle -0.03}$}
\psfrag{0.1}{\hspace*{1em}${\scriptscriptstyle 0.10}$}
\psfrag{0.05}{\hspace*{1em}${\scriptscriptstyle 0.05}$}
\psfrag{-0.05}{\hspace*{.5em}${\scriptscriptstyle -0.05}$}
\psfrag{-0.1}{\hspace*{.5em}${\scriptscriptstyle -0.10}$}
\psfrag{7}{\hspace*{1em}${\scriptscriptstyle 7.0}$}
\psfrag{6.5}{\hspace*{1em}${\scriptscriptstyle 6.5}$}
\psfrag{6}{\hspace*{1em}${\scriptscriptstyle 6.0}$}
\psfrag{5.5}{\hspace*{1em}${\scriptscriptstyle 5.5}$}
\psfrag{order}{$1$}
\psfrag{C1}{\tiny$C_1^{\rm se}$}
\psfrag{C2}{\tiny$C_2^{\rm se}$}
\psfrag{C3}{\tiny$C_3^{\rm se}$}
\begin{center}
  \includegraphics[width=120mm]{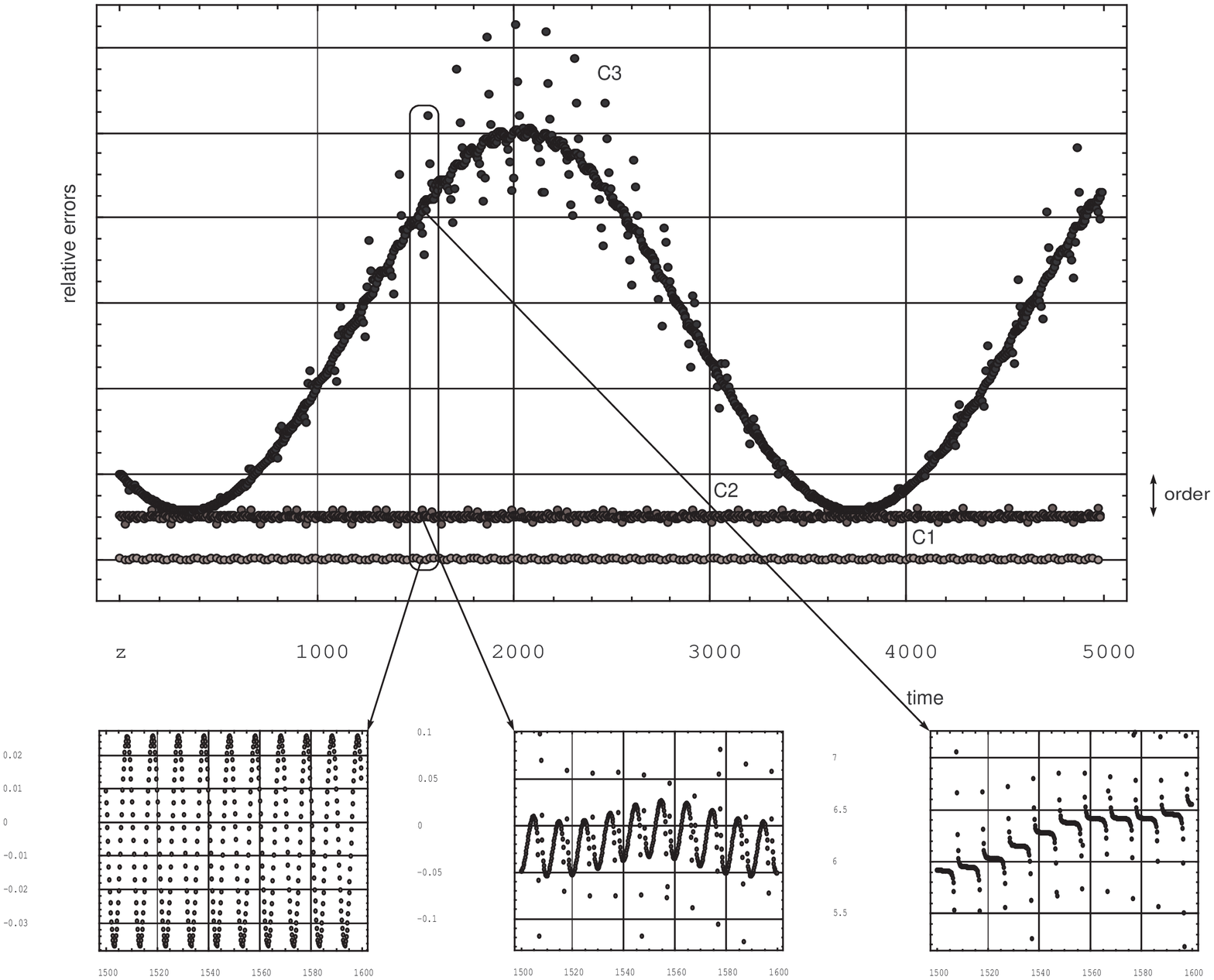}
\end{center}
\caption{The relative errors of the constants of motion,
$C_i^{\rm se}:=C_{i,n}/C_{i,0}$, $n=t/\Delta t$, of 
the discrete solution given by the symplectic Euler 
method~\eref{eq:symplectic_Euler}.}
\label{fig:4}
\end{figure}

Figures~\ref{fig:3} and~\ref{fig:4} present the relative errors 
$C_i^{\rm ec,se}:=C_{i,n}/C_{i,0}$, $n=t/\Delta t$ 
of the discrete solution given by the energy conservation scheme (ec)
and the symplectic Euler method (se)
for the two body case ($N=2$). 
We should note that we do not present all 
the data of each relative errors to keep the size of the data file
reasonable and that the offsets ($=(i-1)\times1.0$) are added 
for convenience of presentation.
The initial condition, the coupling parameter and the strength 
of the harmonic confinement are
the same as those given for the numerical calculation that gives 
figs.~\ref{fig:1} and~\ref{fig:2}.
The discrete time step is set at $\Delta t=0.200$ for these schemes.

One observes that 
the relative errors of the energies ($C_1^{\rm ec}$ and $C_2^{\rm ec}$)
calculated by the energy conservation 
scheme~\eref{eq:energy_conservation_scheme}
remain within the order of the machine epsilon ($=10^{-16}$) times 
the number of iteration ($=t/\Delta t$), which gives a numerical confirmation
that the energies are conserved by the scheme. On the other hand,
one also confirms that the relative errors (except for $C_1^{\rm ec}$ and
$C_2^{\rm se}$) are much larger than the order of machine epsilon times 
the number of iteration, even though one cannot observe growth of 
the relative errors in both two figures. These errors are not brought
about by the round-off errors at the order of machine epsilon times 
the number of iteration, but by the schemes themselves.
One perceives that there is a clear distinction between 
fig.~\ref{fig:2} or the super-integrable 
discretization~\eref{eq:difference_equation} and
figs.~\ref{fig:3} and \ref{fig:4} or the energy 
conservation scheme~\eref{eq:energy_conservation_scheme} and the
symplectic Euler method~\eref{eq:symplectic_Euler}.

\begin{figure}
\psfrag{time}{time $t$}
\psfrag{relative errors}{relative errors}
\psfrag{15}{15}
\psfrag{10}{10}
\psfrag{5}{5}
\psfrag{0}{0}
\psfrag{-5}{$-5$}
\psfrag{-10}{$-10$}
\psfrag{-15}{$-15$}
\psfrag{x1}{$x_1$}
\psfrag{x2}{$x_2$}
\psfrag{super}{\hspace*{-4.5em}super-integrable/analytic}
\psfrag{energy}{\hspace*{-4.3em}energy conservation}
\psfrag{symplectic}{\hspace*{-4.5em}symplectic Euler}
\psfrag{discrete}{\hspace*{-5em}super-integrable discretization}
\psfrag{analytic}{\hspace*{-.5em}analytic solution}
\begin{center}
  \includegraphics[width=120mm]{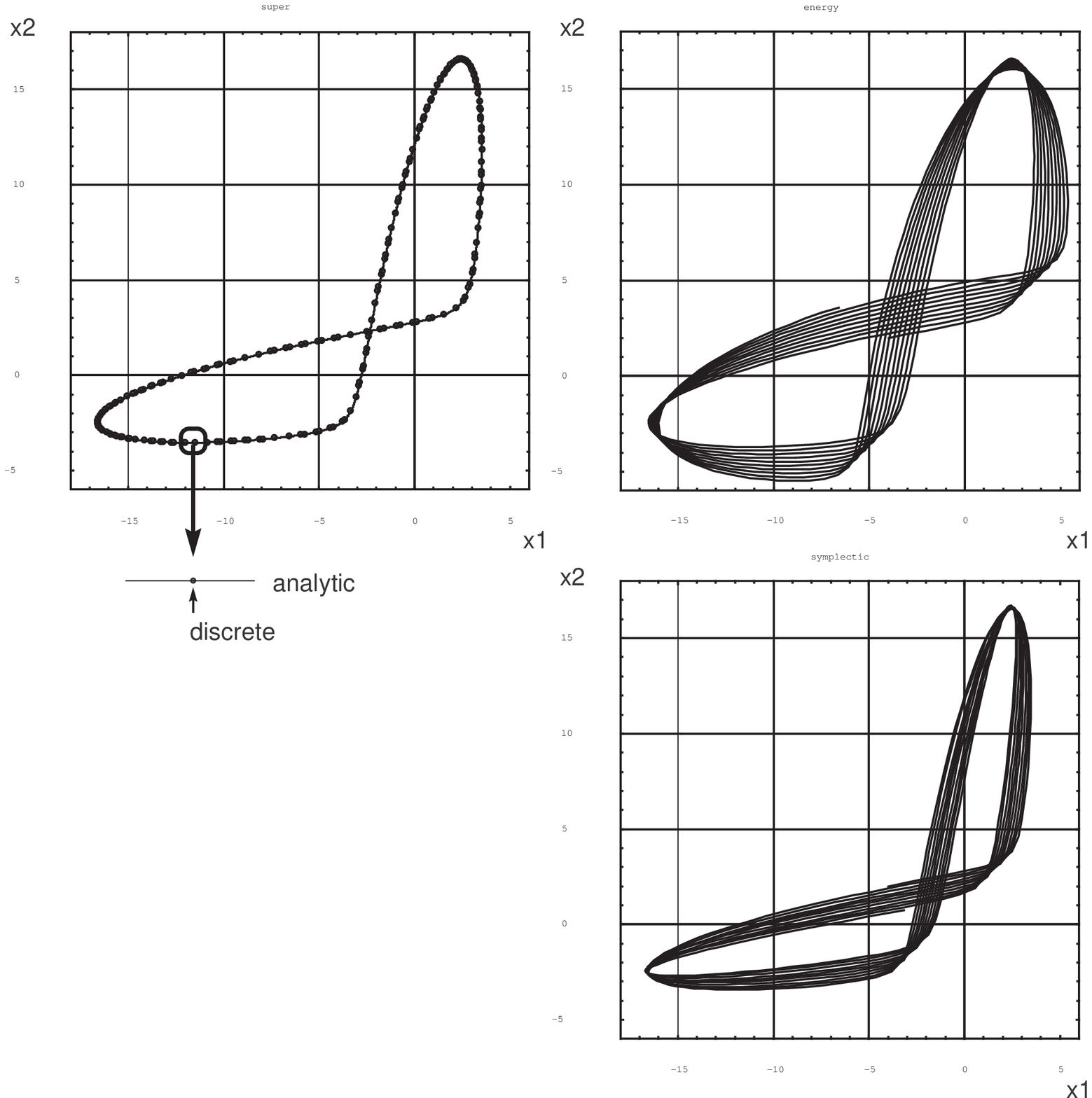}
\end{center}
\caption{The ``Lissajous plots''}
\label{fig:5}
\end{figure}

One of the characteristics of the super-integrable system is that
its bounded orbit is always closed in the phase space and, 
in particular, in the configuration space. As presented 
in fig.~\ref{fig:5}, the
orbit in the $x_1$-$x_2$ plane, or the ``Lissajous plot'' in other words,
thus provides the distinctest way of comparing
the three discretizations.
The time interval of the Lissajous plots is $0\leq t\leq 200$, 
which corresponds to ten periods of motion. While the orbits generated by
the energy conservation scheme and the symplectic Euler method are
not closed, the orbit by the super-integrable discretization is always
on that of the exact analytic solution, which is, of course, closed.

\section{Summary and Concluding Remarks}\label{sec:remarks}

The aim of the paper was to present a time-discretization
for the Calogero model that maintains its super-integrable structure.
As discussed in section~\ref{sec:projection},
the super-integrable structure of the original
Calogero model is provided by the Lax formulation
and the projection method, which are closely related to the equations
of motion of the harmonic oscillator. 
With the help of the integrable discretization of the harmonic 
oscillator~\cite{Hirota2000},
the Lax formulation and the projection method are discretized.
As a consequence,
a time discretization that preserves the
super-integrability of the Calogero model is presented 
in section~\ref{sec:discretization}. In particular for the
two-body case, an explicit form of the difference
equations of motion~\eref{eq:difference_equation} is obtained 
in section~\ref{sec:DEM}. The difference equations give an explicit scheme
for time-integration. Numerical results
by the super-integrable discretization~\eref{eq:difference_equation}
together with comparison with those by the energy conservation 
scheme~\eref{eq:energy_conservation_scheme} 
as well as the symplectic Euler method~\eref{eq:symplectic_Euler}
are presented in five figures in section~\ref{sec:DEM}, which
give a intuitive numerical verification of the super-integrability 
of the scheme.

Lastly, we should give several remarks on previous studies that are 
relevant to the present work. 
In refs.~\cite{Nijhoff1994,Nijhoff1996-1,Nijhoff1996-2} (and
also in ref.~\cite{Suris2003}), integrable discretizations of the
rational Calogero--Moser model (the case $\omega=0$) and its
trigonometric, elliptic and ``relativistic'' (a $q$-difference
generalization with respect to space coordinates) generalizations were 
presented. The structure of the projection method also underlies these
integrable discretizations, but the interaction parameter of the continuous-time
models and the time-step of the discrete-time models are related 
with each other in the integrable discretizations above.
On the other hand, our discretization preserves not only the integrablity
but also the super-integrablity of the Calogero model and both
the time-step $\Delta t$ and the interaction parameter $a$ independently
appear in the discrete equations of motion.
Thus our discretization is apparently different from that of the previous 
studies.
The possibility of the mutual penetration of the particles 
in the time-discrete model was reported in ref.~\cite{Nijhoff1994},
which is certainly unlike the continuous-time case.
But the scheme given there is implicit and it could not privide a way 
to verify this possiblity.
In our super-integrable discretization of the present work, however,
we gave an explicit scheme of the Calogero model for
the two-body case~\eref{eq:difference_equation} 
and numerically observed in fig.~\ref{fig:1} that 
there was no penetration of the particles.
And this observation should be the same for the Calogero--Moser case
corresponding to the limit $\omega\rightarrow 0$ of the present work.
The comparison of the two different discretizations
together with additional consideration on the super-integrability of the
rational Calogero--Moser model~\cite{Wojciechowski1983} will be presented 
in a separate paper. Further studies on the
trigonometric, elliptic and relativistic Calogero--Moser model
along the line of our super-integrable discretization of the Calogero model
as well as to obtain explicit forms of the difference equations of
motion for general $N$-body case (or at least 3-body case)
is worthy of interest.

\section*{Acknowledgements}
Most of this work was carried out during the one-year stay of HU at CRM, 
Universit\'e de Montr\'eal and at McGill University hosted by LV.
HU would like to express his sincere gratitude for their warm hospitality.
This author is also supported by a
grant for his research activities abroad from the Ministry of Education, 
Culture, Sports, Science and Technology of Japan. 
This work initially started in 2001 when HU visited HY at 
National Astronomical Observatory of Japan (NAOJ) with 
a financial support of NAOJ. 
The work of LV is supported in part through a grant from NSERC.
The work of HY is supported in part by the Grant-in-Aid for Scientific
Research of Japan Society for the Promotion of Science (JSPS), No.~15540224.
The authors are grateful to the referee for constructive comments and also
for bringing ref.~\cite{Suris2003} to their attention.

\end{document}